\documentclass[letterpaper,aps,prd,twocolumn,superscriptaddress,tightenlines,showpacs,preprintnumbers,nofootinbib]{revtex4}
\usepackage{graphicx}
\usepackage[intlimits,centertags]{amsmath}
\usepackage{amssymb,amsfonts}
\usepackage{gensymb}
\usepackage{mathrsfs}
\usepackage{hyperref}

\newcommand{\en}{\varepsilon}

\begin{document}

\title{Ensemble Fluctuations of the Flux and Nuclear Composition of\\ Ultra-High Energy Cosmic Ray Nuclei}

\author{Markus~Ahlers} 
\affiliation{Wisconsin IceCube Particle Astrophysics Center (WIPAC) and Department of Physics,\\ University of Wisconsin, Madison, WI 53706, USA\\[0.0cm]}

\author{Luis~A.~Anchordoqui}
\affiliation{Department of Physics, University of
  Wisconsin-Milwaukee, Milwaukee, WI 53201, USA\\[0.0cm]} 

\author{Andrew~M.~Taylor}
\affiliation{Dublin Institute for Advanced Studies, 31 Fitzwilliam Place, Dublin 2, Ireland}

\begin{abstract}
The flux and nuclear composition of ultra-high energy cosmic rays depend on the cosmic distribution of their sources. Data from cosmic ray observatories are yet inconclusive about their exact location or distribution, but provide a measure for the average local density of these emitters. Due to the discreteness of the emitters the flux and nuclear composition is expected to show ensemble fluctuations on top of the statistical variations, {\it i.e.}~``cosmic variance''. This effect is strongest for the most energetic cosmic rays due to the limited propagation distance in the cosmic radiation background and is hence a local phenomenon. For the statistical analysis of cosmic ray emission models it is important to quantify the possible level of this variance. In this paper we present a completely analytic method that describes the variation of the flux and nuclear composition with respect to the local source density. We highlight that proposed future space-based observatories with exposures of ${\cal O} (10^6~{\rm km}^2\,{\rm sr}\,{\rm yr})$ will attain sensitivity to observe these spectral fluctuations in the cosmic ray energy spectrum at Earth relative to the overall power-law fit. 
\end{abstract}


\maketitle

\section{Introduction}

The diffuse spectrum of ultra-high energy (UHE) cosmic rays (CR) is expected to consist of a superposition of fluxes from many individual point-sources distributed throughout the Universe. The absence of significant event clustering across the sky sets a lower limit on the local source density or, equivalently, a lower limit on the number of sources that effectively contribute to the spectrum. Typically this number is very large and the UHE CR spectrum is expected to reflect the average contribution of these sources. It is hence common practice for theoretical studies to approximate the distribution of CR sources via a spatially homogeneous and isotropic emission rate density that reproduces the average source spectrum. This treatment greatly simplifies the study of UHE CRs and can reproduce various spectral features from a simple power-law source spectrum.

The energy spectrum of cosmic rays follows a simple power-law over many energy decades. Small variations of the spectral index can be interpreted either as a transition between CR populations or as an imprint of CR propagation effects. The {\it ankle} -- a hardening of the spectrum at $10^{18.5}$~eV -- could be formed naturally by the superposition of two power-law fluxes and serves as a candidate of the transition between galactic heavy nuclei and extra-galactic cosmic ray protons~\cite{Linsley:1963bk,Hill:1983mk}. It has also been advocated that this feature could be well reproduced by a proton-dominated power-law spectrum, where the ankle is formed as a {\it dip} in the spectrum from the energy loss of protons via Bethe-Heitler pair production~\cite{Hillas:1967,Berezinsky:2002nc}. In this case extra-galactic protons would already have started to dominate the spectrum beyond the {\it 2nd knee}, a feature which corresponds to a slight {\it softening} of the spectrum at $10^{17.7}~{\rm eV}$.

Proton-dominance beyond the ankle is ultimately limited by the onset of photopion production on the cosmic microwave background (CMB), whereas dominance of a heavy composition is restricted by nucleus photodisintegration through the giant dipole resonance (GDR)~\cite{Greisen:1966jv,Zatsepin:1966jv} -- the so called {\it Greisen-Zatsepin-Kuz'min} (GZK) suppression  at around $10^{19.7}$~eV. Indeed, a flux suppression in this energy region has been observed in the HiRes and Auger data with a high statistical significance~\cite{Abbasi:2007sv,Abraham:2008ru,Abraham:2010mj}. As noted elsewhere~\cite{Ahlers:2009rf,Hooper:2010ze}, secondary neutrinos and $\gamma$-rays of these hadronic interactions can serve as additional discriminators between various CR models.

In order to collect the elusive events above $10^{19.7}~{\rm eV}$ (which present an integrated flux of less than 1 event per km$^{2}$ per steradian and per century) observatories with large apertures and long exposure time are needed. Today, the leading role in CR is played by ground based facilities that cover vast areas with particle detectors overlooked by fluorescence telescopes. The largest is the Pierre Auger Observatory, with a surface detector array of $1600$ water Cherenkov tanks covering $3000$~km$^{2}$, which accumulates annually about $6 \times 10^3~{\rm km}^2 \, {\rm sr} \, {\rm yr}$ of exposure~\cite{Abraham:2010zz}. The more recently constructed Telescope Array (TA) covers 700~km$^{2}$ with 507 scintillator detectors~\cite{AbuZayyad:2012kk}, which should accumulate annually about $1.4 \times 10^3~{\rm km}^2 \, {\rm sr} \, {\rm yr}$ of exposure. 

In the near future, the JEM-EUSO mission will orbit the Earth on board the International Space Station at an altitude of of about 400~km. Whilst in the ``nadir'' mode, the remote-sensing space instrument (with $\pm 30^\circ$ field of view) will monitor an area of approximately $1.3 \times 10^5~{\rm km}^{2}$, recording video clips of fast UV flashes by sensing the fluorescence light produced through charged particle interactions. This innovative pathfinder  mission will observe approximately $6 \times 10^4~{\rm km}^2 \, {\rm sr} \, {\rm yr}$ annually~\cite{Adams:2012hr}, a factor of 10 above Auger.

In this paper we elaborate on the question as to what extent the spectral information in the GZK region can be used to discriminate between different CR source composition models. Due to the strength of the GZK mechanism the spectrum in this region is dominated by (and requires the presence of) local sources~\cite{Taylor:2011ta}. In this case the flux from a few CR sources can significantly fluctuate from a homogeneous distribution that is typically assumed in CR flux predictions~\cite{MiraldaEscude:1996kf,Anchordoqui:1997rn,MedinaTanco:1998uh,Aloisio:2010wv}. In contrast to Poisson fluctuations in the GZK region~\cite{DeMarco:2003ig} the manifestations of ensemble fluctuations persist in the limit of large event statistics. We will quantify these stochastic fluctuations in the following utilizing an analytic solution to the flux of CR nuclei derived in Refs.~\cite{Hooper:2008pm,Ahlers:2010ty}. 

We will start in Sec.~\ref{sec:propagation} with a brief review of the propagation of CR nuclei and the calculation of the mean observed fluxes. In Sec.~\ref{sec:variation} these results will be used to derive an analytic approximation of the flux and mean mass variations due to the distribution of sources. We summarize our findings in Sec.~\ref{sec:conclusion}.

\section{Propagation of Cosmic Ray Nuclei}\label{sec:propagation}

Since cosmic rays are subject to deflections in galactic and inter-galactic magnetic fields the observed CR events do not point directly back to their sources. The identification of the CR sources is hence experimentally challenging and has so far proved inconclusive. There is a general consensus that the sources which are responsible of the UHE CR spectrum are of extra-galactic origin. These sources are expected to follow a spatially homogeneous distribution and the mean (ensemble-averaged) flux of UHE CRs (of type $i$) follows a set of (Boltzmann) continuity equations of the form:
\begin{multline}\label{eq:diff0}
\dot Y_i = \partial_E(HEY_i) + \partial_E(b_iY_i)-\Gamma^{\rm tot}_{i}\,Y_i
\\+\sum_j\int{\rm d} E_j\,\gamma_{ji}Y_j+\mathcal{H}\mathcal{Q}_i\,,
\end{multline}
together with the Friedman-Lema\^{\i}tre equations describing the cosmic expansion rate $H(z)$ as a function of red-shift $z$. We follow the usual cosmological {\it concordance model} dominated by a cosmological constant with $\Omega_{\Lambda} \sim 0.7$ and a (cold) matter component, $\Omega_{\rm m} \sim 0.3$ where \mbox{$H^2 (z) = H^2_0\,[\Omega_{\rm m}(1 + z)^3 + \Omega_{\Lambda}]$}, normalized to its value today of $H_0 \sim70$ km\,s$^{-1}$\,Mpc$^{-1}$~\cite{Beringer:1900zz}. The time-dependence of the red-shift can be expressed via ${\rm d}z = -{\rm d} t\,(1+z)H$.  The first and second terms on the r.h.s.~of Eq.~(\ref{eq:diff0}) describe, respectively, red-shift and other continuous energy losses (CEL) with rate $b \equiv \mathrm{d}E/\mathrm{d}t$.  The third and fourth terms describe more general interactions involving particle losses ($i \to$ anything) with total interaction rate $\Gamma^{\rm tot}_i$, and particle generation of the form $j\to i$ with differential interaction rate $\gamma_{ij}$.  The last term on the r.h.s.~corresponds to the emission rate of CRs of type $i$ per co-moving volume, depending on the emission rate $\mathcal{Q}_i$ per source and their density $\mathcal{H}$.  

The two main reactions of UHE CR nuclei during their cosmic evolution are
photodisintegration~\cite{Stecker:1969fw,Puget:1976nz,Epele:1998ia,RachenTHESIS,Stecker:1998ib,Khan:2004nd,Allard:2005ha,Goriely:2008zu} and Bethe-Heitler pair production~\cite{Blumenthal:1970nn} with the cosmic radiation background. In addition to the dominant contribution of the CMB we also include the infra-red/optical background from Ref.~\cite{Franceschini:2008tp} in our calculation of interaction and energy loss rates. Photodisintegration is dominated by the GDR with main branches $A\to(A-1)+N$ and $A\to(A-2)+2N$ where $N$ indicates a proton or neutron~\cite{Puget:1976nz}. The GDR peak in the rest frame of the nucleus lies at at about $20$~MeV for one-nucleon emission, corresponding to $E^A_{\rm GDR} \simeq A\times 2\times\epsilon^{-1}_{\rm meV}\times10^{19}$~eV in the cosmic frame with photon energies $\epsilon = \epsilon_{\rm meV}$~meV. At energies below 10~MeV there exist typically a number of discrete excitation levels that can become significant for low mass nuclei. Above 30~MeV, where the photon wavelength becomes comparable or smaller than the size of the nucleus, the photon interacts via substructures of the nucleus. Out of these the interaction with quasi-deuterons is typically most dominant and forms a plateau of the cross section up to the photopion production threshold at $\sim145$~MeV. Bethe-Heitler pair production can be treated as a continuous energy loss process with rate $b_A(z,E) = Z^2b_p(z,E/A)$, where $b_p$ is the energy loss rate of protons~\cite{Blumenthal:1970nn}. The (differential) photodisintegration rate $\Gamma_{A\to B}(E)$ ($\gamma_{A\to B}(E,E')$) is discussed in more detail in Ref.~\cite{Ahlers:2010ty}.

The evolution of the spectra proceeds very rapidly on cosmic time scales and the diffuse flux of secondary nuclei, $J$, looks generally quite different from the initial source injection spectrum~\cite{Anchordoqui:1997rn}. The reaction network of nuclei depend in general on a large number of stable or long-lived isotopes. If the life-time of an isotope is much shorter than its photodisintegration rate it can be effectively replaced by its long-lived decay products in the network~(\ref{eq:diff0}). Typically, neutron-rich isotopes $\beta$-decay to a stable or long-lived nucleus with the same mass number. In most cases there is only one stable nucleus per mass number below ${}^{56}$Fe with the exception of the pairs ${}^{54}$Cr/${}^{54}$Fe, ${}^{46}$Ca/${}^{46}$Ti, ${}^{40}$Ar/${}^{40}$Ca and ${}^{36}$S/${}^{36}$Ar. We follow here the approach of Puget, Stecker and Bredekamp (PSB)~\cite{Puget:1976nz} and consider only a single nucleus per mass number $A$ in the decay chain of primary iron ${}^{56}$Fe. This PSB-chain of nuclei linked by one-nucleon losses is discussed in more detail in Ref.~\cite{Ahlers:2010ty}.

Note that the set of Boltzmann Eqs.~(\ref{eq:diff0}) does not take into account the deflection of charged CR nuclei during their propagation through magnetic fields. Magnetic scattering can be viewed as a diffusion process that depends on the particle's Larmor radius $R_{\rm L} = E/(ZeB) \simeq{1.1}{\rm Mpc}\,E_{\rm EeV}/(ZB_{\rm nG})$ and the coherence length of the magnetic field. At energies where the diffusion length becomes larger than the distance to the nearest CR source the CR spectrum is expected to be suppressed. It has been speculated that for particularly strong inter-galactic magnetic fields of strength $\sim 1$~nG and coherence length of $\sim 1$ Mpc (see {\it e.g.}~Ref.~\cite{Kronberg:1993vk}), the diffusive propagation of CR protons can start to affect the spectrum below about $10^{18}$~eV~\cite{Aloisio:2004fz}. For heavier nuclei diffusive propagation can in principle remain important up to higher energies due to the dependence $R_\mathrm{\rm L} \propto 1/Z$~\cite{Harari:1999it}. The results of this paper assume that the contribution of inter-galactic or galactic magnetic fields can be neglected for the calculation of the UHE CR spectrum.

\begin{figure}[t]\centering
\includegraphics[width=0.95\linewidth]{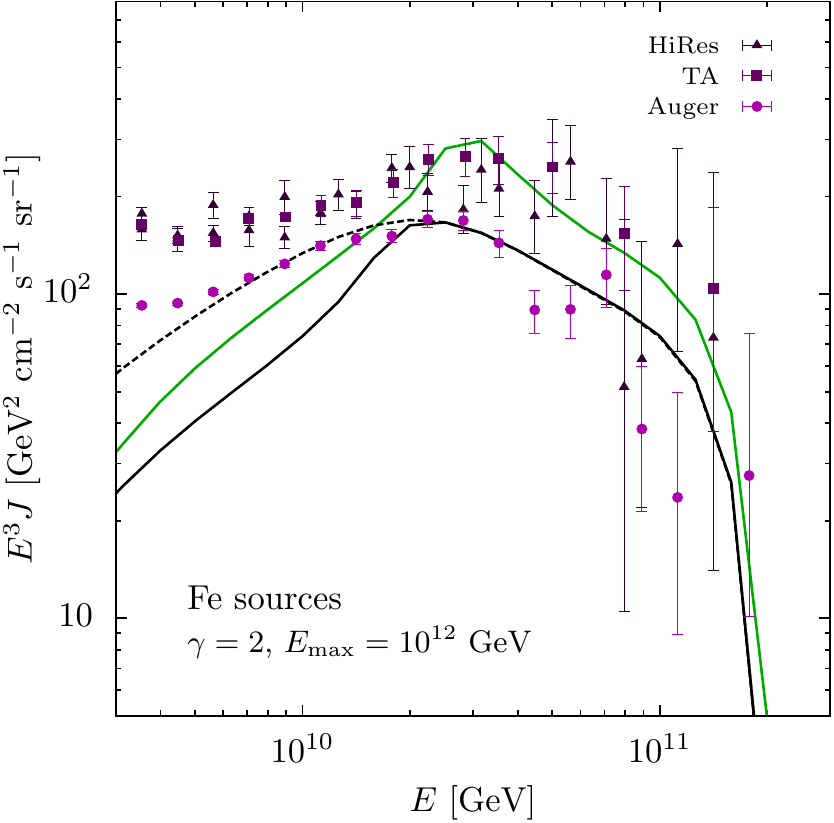}
\caption[]{The spectra from pure iron sources with exponential cutoff $E_{\rm max}=10^{12}$~GeV and power index $\gamma=2$. The upper solid (green) line shows the result of the analytic approximation (\ref{eq:meanN}) for a homogeneous distribution of sources $r_{\rm min}<r<r_{\rm max}$ with $r_{\rm min}=10$~Mpc and $r_{\rm max}=1$~Gpc. The lower solid (black) line is the corresponding numerical calculation based on Eqs.~(\ref{eq:diff0}) including redshift scaling of interaction rates and energies. The relative difference between these calculations can be traced back to the onset of redshift scaling for propagation distances of the order of Gpc (see main text). The dashed black line shows a model including sources up to $z_{\rm max}=2$. Also shown is recent data from HiRes~\cite{Abbasi:2007sv}, Auger~\cite{Abraham:2010mj} and the Telescope Array (TA)~\cite{AbuZayyad:2012ru}.}\label{fig1}
\end{figure}

As an example we show in Fig.~\ref{fig1} the solution of Eq.~(\ref{eq:diff0}) for an iron source model using a power-law spectral emission rate $Q_{\rm Fe} \propto E^{-\gamma}\exp(-E/E_{\rm max})$ with $\gamma=2$ and $E_{\rm max}=10^{21}$~eV. This model is motivated by a previous study~\cite{Taylor:2011ta} and reproduces the Auger data above the ankle within systematic uncertainties. The dashed black line corresponds to source contributions above $r_{\rm min} =10$~Mpc extending up to redshift $z_{\rm max}=2$ where no red-shift evolution of the emission is assumed, {\it i.e.}~$\mathcal{H}\propto1$. The solid black line marks the local contribution up to 1~Gpc calculated with the same method. This local contribution where red-shift scaling of energies and interaction rates is suppressed can be approximated by an analytic solution which is shown as the green solid line in the plot. We will discuss this method in the next section.

\section{Ensemble Average and Variation}\label{sec:variation}

Photodisintegration and photopion interactions of nuclei happen on timescales much shorter than the Hubble scale. Since we are interested in the variation of the average flux and mass composition from the distribution of local CR sources we will neglect redshift scalings in the following. In this case the Green's function of the Boltzmann equations~(\ref{eq:diff0}) can be expressed in a simple analytic form, as discussed in Refs.~\cite{Hooper:2008pm,Ahlers:2010ty}. 

It is convenient to study the flux of nuclei with mass number $A$ in terms of the energy per {\it nucleon} $\en=E/A$ that is conserved under photodisintegration. Introducing the CR flux $F_{A,i} \equiv \Delta \en_iA\,{{\rm d}F}_{A}(A\en_i)/{{\rm d} E}$ per nucleon energy bin $i$ with bin-width $\Delta\en_i$ (centered at a nucleon energy $\en_i$) and corresponding emission rates $Q_{A,i} \equiv \Delta \en_i A \mathcal{Q}_A(A\en_i)$ we find an analytic solution of the form
\begin{equation}\label{eq:fullsol}
F_{A,i}(r) \simeq \sum_{\mathbf{c}}\sum_{k=1}^{n_c}\mathcal{A}_k(\mathbf{c})\frac{e^{-r\Gamma^{\rm tot}_{c_k}}}{4 \pi r^2}Q_{c_1}\,.
\end{equation}
The sum on the r.h.s.~runs over all possible production paths $\mathbf{c}$ of a CR nucleus with final mass number $A$ and nucleon energy bin $i$. Each of the $n_c$ elements $c_k$ of the path $\mathbf{c}$ consists of a doublet $(B,j)$ denoting the mass number $B$ and nucleon energy bin $j$ of intermediate nuclei. The first element $c_1$ of the paths corresponds to the nucleus emitted from a source at a rate $Q_{c_1}$ (possibly equal to zero depending on source composition) and the last element is fixed at $c_{n_c} = (A,i)$. Each path is weighted by a set of $n_c$ dimensionless amplitudes $\mathcal{A}_k(\mathbf{c})$ that are independent of the source distance $r$. For more details we refer to Ref.~\cite{Ahlers:2010ty} and Appendix~\ref{sec:analytic}.

We now want to study the statistical mean and variation of the aggregated flux of $n_s$ local CR sources denoted by 
\begin{equation}\label{eq:NAi}
N_{A,i}\equiv\sum\limits_{s=1}^{n_s}F_{A,i}(r_s)\,,
\end{equation}
as well as its corresponding mass composition. Herein we assume that the probability distribution function (PDF) for local ($r/H_0\ll1$) sources is flat in Euclidean space. Let $n_s$ sources be distributed between redshift $r_{\rm min}$ and $r_{\rm max}$. The number of sources can then be expressed via the (local) source density $\mathcal{H}_0$ as $n_s = \mathcal{H}_0 (4\pi/3)(r_{\rm max}^3-r_{\rm min}^3)$. The PDF of a single source is then given by
\begin{equation}\label{eq:prob1}
p(r) = \frac{\mathcal{H}_0}{n_s}4\pi r^2\Theta(r-r_{\rm min})\Theta(r_{\rm max}-r)\,. 
\end{equation}
Note that the local density $\mathcal{H}_0$ is limited by auto-correlation studies of UHE CR events; the lower $\mathcal{H}_0$ the larger the average emission rate of the sources and the greater the chance of local event clusters across the sky. A local source density of $\mathcal{H}_0=10^{-5}$~Mpc${}^{-3}$ is consistent with the absence of ``repeaters'' in CR data~\cite{Waxman:1996hp,Kashti:2008bw}. Values as low as $\mathcal{H}_0=10^{-6}$~Mpc${}^{-3}$ are still marginally consistent with auto-correlation studies of UHE CR nuclei~\cite{Takami:2012uw}. We will consider these two cases as fiducial values in the following. 

Following Ref.~\cite{Lee:1979} the ensemble-average of a quantity $X(r_1,\ldots,r_{n_s})$ depending on the distance of the $n_s$ sources can then be expressed as
\begin{equation}
\langle X\rangle = \int {\rm d}r_1\cdot\ldots\cdot{\rm d}r_{n_s}p(r_1)\cdot\ldots\cdot p(r_{n_s})X\,.
\end{equation}
The ensemble-average of the local flux of CR nuclei using Eqs.~(\ref{eq:NAi}) and (\ref{eq:fullsol}) is then simply
\begin{gather}
\langle N_{A,i}\rangle \equiv \mathcal{H}_0\int_{r_{\rm min}}^{r_{\rm max}} {\rm d}r'4\pi r'^2 F_{A,i}(r')\,.
\end{gather}
Using the abbreviation (\ref{eq:A}) we can then write this in an analytic form as
\begin{equation}\label{eq:mean}
\langle N_{A,i}\rangle=\sum_{\mathbf{c}}\sum_{k=1}^{n_c}\mathcal{A}_k(\mathbf{c})\xi(\Gamma^{\rm tot}_{{c}_{k}})Q_{c_{1}}\,,
\end{equation}
where we define
\begin{equation}\label{eq:xi}
\xi(\Gamma) \equiv \frac{\mathcal{H}_0}{\Gamma}\left(e^{-r_{\rm min}\Gamma}-e^{-r_{\rm max}\Gamma}\right)\,.
\end{equation}

From the experimental point of view the interesting quantities are the ensemble-averaged total flux of nuclei $N_{\rm tot}$ and mean mass number $A_{\rm av}$ at the highest CR energies $E$. The mean total flux is simply given by Eq.~(\ref{eq:mean}) as\footnote{Here and in the following $N_{A}(E/A)$ is a short-hand notation for $N_{A,i}$ with $\en_i=E/A$.}
\begin{equation}\label{eq:meanN}
\langle N_{\rm tot}(E)\rangle \equiv \sum_A \langle N_{A}(E/A)\rangle\,.
\end{equation}
We now return to the example of an iron source model shown in Fig.~\ref{fig1}. The solid green line in Fig.~\ref{fig1} shows the result of the analytic approximation~(\ref{eq:mean}). This approximation agrees with the numerical result (solid  black line) within a factor two or better depending on the energy. The relative difference is expected from the onset of redshift scaling for propagation distances of the order of Gpc ($z\simeq1/4$). At low energies this introduces a relative upward shift of the flux of $\simeq30$\% for a source model with $\gamma=2$ and $\mathcal{H}\propto1$. This agrees well with the result of the calculations. In addition, threshold effects that lead to breaks in the spectrum scale with redshift as $E_{\rm th}\propto 1/(1+z)^2$ and are shifted to lower energies by up to $\simeq50$\%. This effect can also be noticed by the relative position of the break in Fig.~\ref{fig1}.

\begin{figure*}[p]\centering
\includegraphics[width=0.82\linewidth]{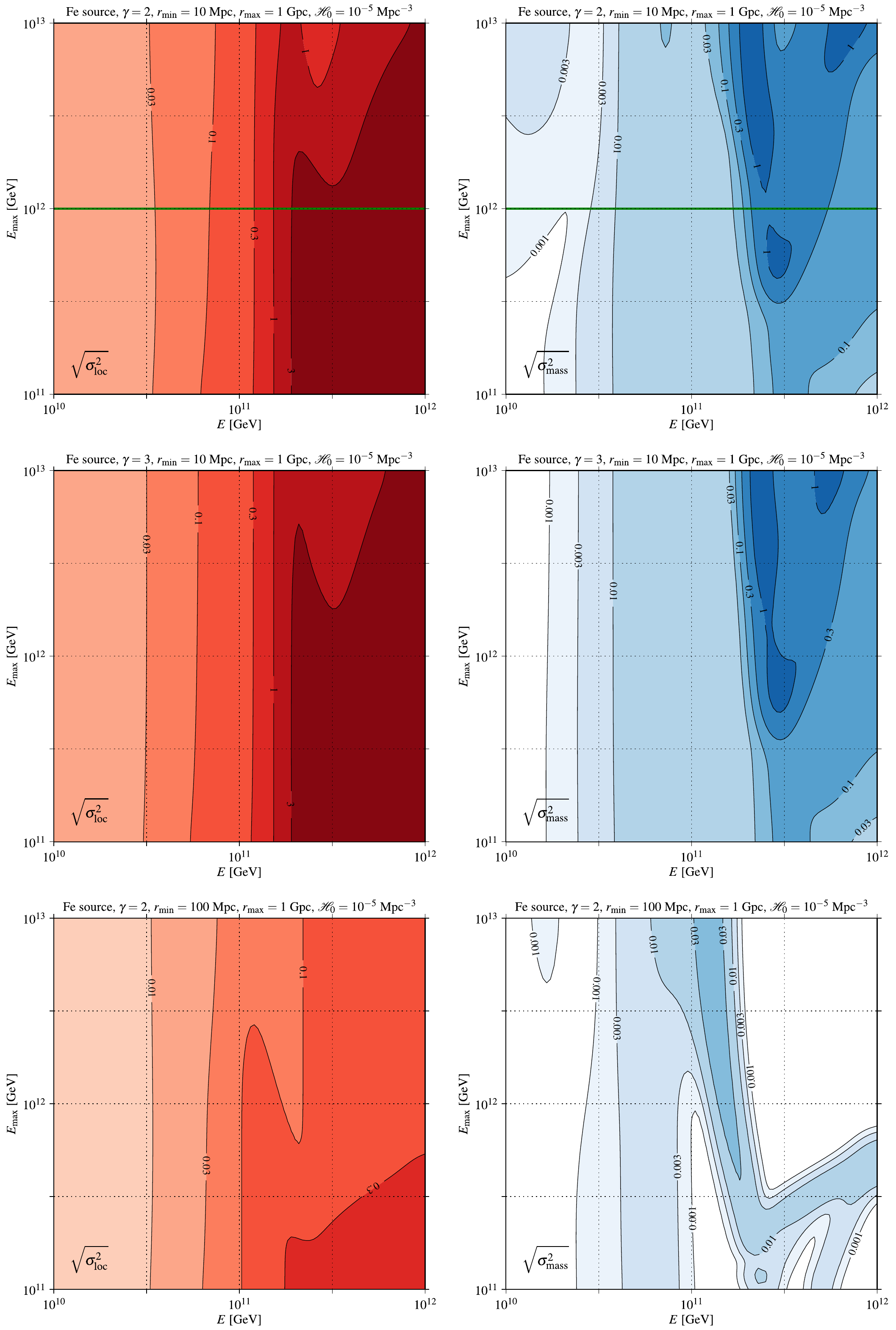}
\caption[]{The local relative error of the flux (Eq.~(\ref{eq:locsig}); left plots in red) and average mass composition (Eq.~(\ref{eq:sigA}); right plots in blue) for a distribution of iron sources with the model parameters indicated above the plots. We show contour plots in terms of the observed CR energy $E$ of the iron nucleus and the exponential cutoff $E_{\rm max}$ of the emission. The solid (green) line in the top plots indicate the corresponding relative error for the model shown in Fig.~\ref{fig1}. All calculations assume a local source density of $\mathcal{H}_0=10^{-5}$~Mpc${}^{-3}$ and scale as $\mathcal{H}_0^{-1/2}$.}\label{fig2}
\end{figure*}

This example illustrates the limitations of the analytic solution for the calculation of large scale (early-time) contributions to the CR flux. However, the analytic approximation provides a convenient description of the nuclei cascades that happen on small scales and depend on the local source distribution. In particular, it enables us to study ensemble-variations of the flux. Defining the residual $\delta X = X-\langle X\rangle$, {\it etc.}, we have as usual $\langle\delta X\delta Y\rangle = \langle XY\rangle-\langle X\rangle\langle Y\rangle$. Using Eqs.~(\ref{eq:fullsol}) and (\ref{eq:NAi}) we can then write the variation of the CR flux in an explicit analytic form as
\begin{multline}\label{eq:var}
\langle\delta N_{A,i} \delta N_{B,j}\rangle\\
=\sum_{\mathbf{c},\bar{\mathbf{c}}}\sum_{k=1}^{n_c}\sum_{\bar k=1}^{n_{\bar c}}\mathcal{A}_k(\mathbf{c})\mathcal{A}_{\bar k}(\bar{ \mathbf{c}})
\zeta(\Gamma^{\rm tot}_{{c}_{k}}+\Gamma^{\rm tot}_{{\bar c}_{\bar k}})Q_{c_1}Q_{\bar c_1}\\-\frac{1}{n_s}\langle N_{A,i}\rangle\langle N_{B,j}\rangle\,,
\end{multline}
where $\zeta$ is defined as the expression
\begin{equation}
\zeta(\Gamma) \equiv \mathcal{H}_0\int_{r_{\rm min}}^{r_{\rm max}}{\rm d}r\frac{e^{-r\Gamma}}{4\pi r^2}
\,\,.\label{eq:zeta}
\end{equation}
Note that the last term in Eq.~(\ref{eq:var}) is sometimes omitted since the number of sources $n_s$ is expected to be large, but we keep it in our calculations.
Based on these definitions we can express the relative variation of the total flux via the two-point density perturbations (\ref{eq:var}) as 
\begin{equation}\label{eq:locsig}
\sigma^2_{\rm loc}(E) = \sum_{A,B}\frac{\langle\delta N_{A}(E/A) \delta N_{B}(E/B)\rangle}{\langle N_{\rm tot}(E)\rangle^2}
\,.
\end{equation}

With Eq.~(\ref{eq:meanN}) we can also {\it define} the mean mass number as
\begin{equation}\label{eq:meanA}
\langle A_{\rm av}(E)\rangle \equiv \sum_AA\frac{\langle N_{A}(E/A)\rangle}{\langle N_{\rm tot}(E)\rangle}\,.
\end{equation}
Note  that Eq.~(\ref{eq:meanA}) is in the strict sense not the ensemble-average but serves as a first order estimator. For small fluctuations around the mean value we can approximate the relative variation of the mean mass number (\ref{eq:meanA}) via the two-point correlation function (\ref{eq:var}) as
\begin{multline}\label{eq:sigA}
\sigma^2_{\rm mass}(E) \simeq \sum_{B,C}\frac{\langle\delta N_{B}(E/B) \delta N_{C}(E/C)\rangle }{\langle N_{\rm tot}(E)\rangle^2}\\
\times\left(1-\frac{B}{\langle A_{\rm av}(E)\rangle}\right)\left(1-\frac{C}{\langle A_{\rm av}(E)\rangle}\right)\,.
\end{multline}

In Figure~\ref{fig2} we show contour plots of the relative error of the flux (red; left plots) and mass composition (blue; right plots) for the case of iron sources with different model parameters indicated above the plots. The axes show the observed CR energy vs.~the exponential cutoff energy $E_{\rm max}$. For the calculation we introduced logarithmic energy bins of $\Delta \log_{10}E/{\rm GeV} = 0.02$ and smoothed the result with a Gaussian kernel to account for an experimental resolution of $0.1$. This procedure smoothes out features in the relative variance that are beyond the experimental resolution and result from the rapid nuclei transitions in the GZK region in combination with the relative CR energy shift with the mass of the daughter nuclei.

The results do not strongly depend on the spectral index as can be seen for the cases $\gamma=2$ and $\gamma=3$ with otherwise equal parameters that are shown in the plots of the first two rows. This set of plots assumes a local distribution between 10~Mpc and 1~Gpc. The relative errors are significantly reduced as we increase the distance to the closest source to 100~Mpc as shown in the plots of the last row in Fig.~\ref{fig2}. However, this is only marginally reproducing the UHE CR spectrum as pointed out in Ref.~\cite{Taylor:2011ta}. Note that the green lines in the plots of the top row mark the contribution for the iron source model with $E_{\rm max}=10^{21}$~eV considered in Fig.~\ref{fig1}.

One can also notice from the lower plots of Fig.~\ref{fig2} that the relative error of the average mass composition is below 1\% for CR energies below $10^{19.5}$~eV. This energy marks the end of the energy region where CR event statistics allow an inference of the mass composition from CR data. Note that all the plots in Fig.~\ref{fig2} show the case of a local source density of $\mathcal{H}_0=10^{-5}$~Mpc${}^{-3}$ and levels increase as $\mathcal{H}_0^{-1/2}$. Hence, even for a density $\mathcal{H}_0=10^{-6}$~Mpc${}^{-3}$, still marginally consistent with the data, the ensemble fluctuation on the average mass composition probed by present generation experiments may be safely neglected.

So far we have considered the case that $\Gamma /H_0 \ll 1$ valid at the highest CR energies where we can neglect cosmological contributions of the sources and treat the problem as effectively local. In the opposite case, $\Gamma /H_0 \gg 1$, it is also possible to give an analytic expression of the statistical variation of the CR flux. We assume that $n_s$ sources are isotropically distributed between redshift $z_{\rm min}$ and $z_{\rm max}$ with comoving density $\mathcal{H}(z)$, and local density $\mathcal{H}_0$. The number of sources is then  given by
\begin{equation}\label{eq:ns}
n_s = \int{\rm d}z\frac{{\rm d}\mathcal{V}}{{\rm d}z }\mathcal{H}(z) = \int \frac{{\rm d}z}{H(z)}4\pi d^2_C(z)\mathcal{H}(z)
\end{equation}
where the comoving volume is $\mathcal{V}_C(z) = (4\pi/3) d^3_C(z)$ with comoving distance (in a flat universe) $d_C(z) = \int_0^z{\rm d} z'/H(z')$. The PDF of a single source as a function of redshift is then a simple generalization of Eq.~(\ref{eq:prob1}),
\begin{equation}\label{eq:prob2}
p(z) = \frac{1}{H(z)}\frac{\mathcal{H}(z)}{n_s}4\pi d^2_C(z)\,. 
\end{equation}
As long as only adiabatic redshift scaling is involved the flux of a single source is given by ${\rm d}F/{\rm d} E = \mathcal{Q}((1+z)E)/(4\pi d_C^2)$.
Assuming $\mathcal{Q}(E) \propto E^{-\gamma}$ we then obtain a relative variation of
\begin{equation}\label{eq:adisig}
\sigma_{\rm adi}^2 = \frac{\int{\rm d} z\,p(z)\left[(1+z)^{-\gamma}/d^2_C(z)\right]^2}{\left[\int{\rm d} z\,p(z)(1+z)^{-\gamma}/d^2_C(z)\right]^2}-\frac{1}{n_s}\,.
\end{equation}
We show values for $n_s$ and $(\sigma^2_{\rm adi})^{1/2}$ in Table~\ref{tab1}. Obviously, the statistical variation of the average mass number is negligible in this regime.

\begin{table}[t]
\renewcommand{\arraystretch}{1.2}
\begin{tabular}{c|c|c|c|c|c}\hline \hline
\begin{minipage}[c][0.5cm][c]{1cm}$n$\end{minipage}&
\begin{minipage}[c][0.5cm][c]{1cm}$z_{\rm min}$\end{minipage}&
\begin{minipage}[c][0.5cm][c]{1cm}$z_{\rm max}$\end{minipage}&
\begin{minipage}[c][0.5cm][c]{1cm}$\gamma$\end{minipage}&
\begin{minipage}[c][0.5cm][c]{1.5cm}$n_s$ [$10^6$]\end{minipage}&
\begin{minipage}[c][0.5cm][c]{1.5cm}$\sqrt{\sigma^2_{\rm adi}}$ [\%]\end{minipage}\\\hline 
$0$&$0.01$&$2$&$2$&$5.3$&$0.65$\\
$3$&$0.01$&$2$&$2$&$73$&$0.15$\\
$0$&$0.1$&$2$&$2$&$5.3$&$0.17$\\
$3$&$0.1$&$2$&$2$&$73$&$0.04$\\
$0$&$0.01$&$2$&$2.5$&$5.3$&$0.76$\\
$3$&$0.01$&$2$&$2.5$&$73$&$0.20$\\
$0$&$0.25$&$2$&$2$&$5.2$&$0.09$\\
$3$&$0.25$&$2$&$2$&$73$&$0.02$\\\hline
\multicolumn{3}{c|}{SFR ($z_{\rm min}=0.01$)} &$2$&$173$&$0.14$\\
\multicolumn{3}{c|}{SFR ($z_{\rm min}=0.01$)}&$2.5$&$173$&$0.19$\\\hline
\hline
\end{tabular}
\caption[]{Estimated source number and adiabatic variation defined in Eqs.~(\ref{eq:ns}) and (\ref{eq:adisig}). We show results for a power-law cosmological evolution as $\mathcal{H}= \mathcal{H}_0(1+z)^n$ with $z_{\rm min}<z<z_{\rm max}$ and for the star formation rate (SFR) according to Ref.~\cite{Hopkins:2006bw,Yuksel:2008cu} (with cutoff $z_{\rm min}=0.01$). In all cases we assume a local density of $\mathcal{H}_0=10^{-5}$~Mpc${}^{-3}$.}\label{tab1}
\end{table}

\begin{figure*}[t]\centering
\includegraphics[width=0.45\linewidth]{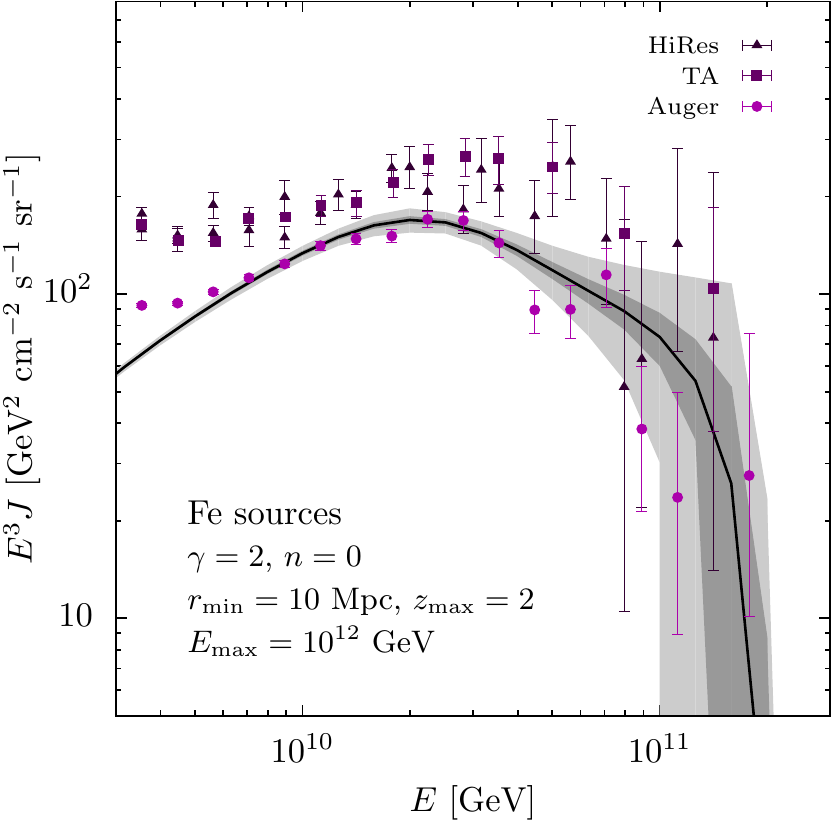}\hspace{0.8cm}\includegraphics[width=0.45\linewidth]{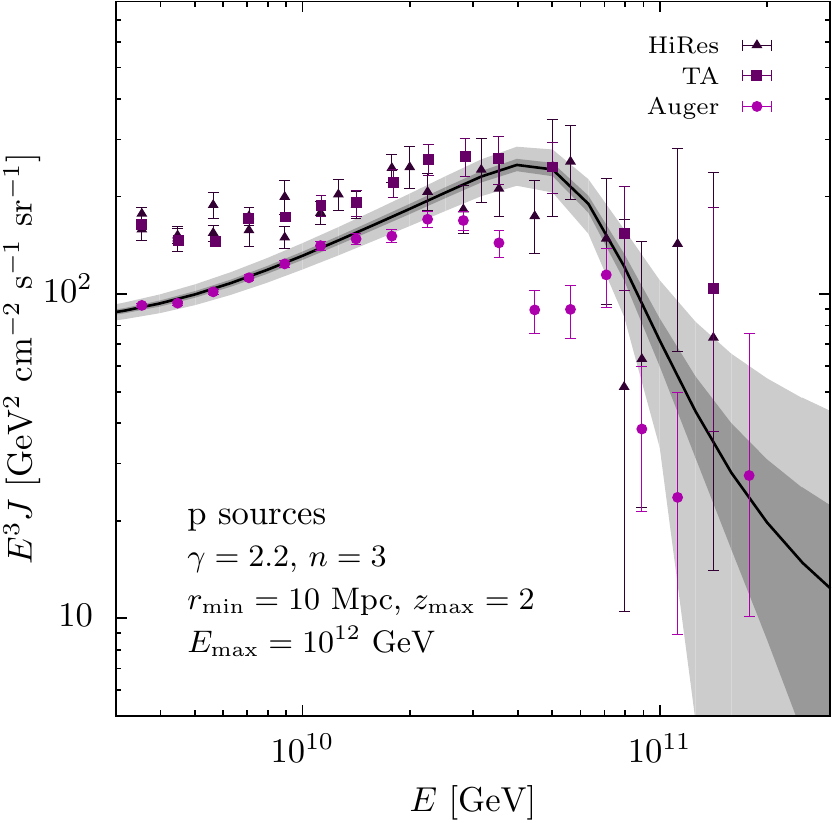}
\caption[]{{\bf Left panel:} The example of Fig.~\ref{fig1} including the approximate variation of the flux assuming a local source distribution $\mathcal{H}_0=10^{-5}$~Mpc${}^{-3}$ (dark gray band) and $\mathcal{H}_0=10^{-6}$~Mpc${}^{-3}$ (light gray band). {\bf Right panel:} For comparison, a proton model with similar source parameters.}\label{fig3}
\end{figure*}

The two limiting behaviors of the relative flux variation (\ref{eq:locsig}) and (\ref{eq:adisig}) motivates the following treatment of the overall relative flux variation. Using the numerical result of the set of Boltzmann Eqs.~(\ref{eq:diff0}) we can calculate the average flux for the local ($r\lesssim1$~Gpc), and global source distribution. This corresponds to the solid and dashed lines, respectively, in Fig.~\ref{fig1}. The flux variation can then be approximated via the superposition
\begin{equation}\label{eq:combinedsig}
\sigma_N^2(E) \simeq (1-x)^2\sigma^2_{\rm adi} + x^2\sigma^2_{\rm loc}(E)\,,
\end{equation}
where $\sigma^2_{\rm loc}$ is calculated from the analytic approximation (\ref{eq:locsig}), $\sigma^2_{\rm adi}$ the adiabatic limit (\ref{eq:adisig}) for $r_{\rm min}\simeq1$~Gpc ($z_{\rm min}\simeq0.25$) and $x=\langle N_{\rm loc}(E)\rangle/\langle N_{\rm global}(E)\rangle$ from the numerical evaluation.

In the left plot of Fig.~\ref{fig3} we show the result of this procedure for the previous example of iron sources shown in Fig.~\ref{fig1}. The solid line (corresponding to the dashed line in Fig.~\ref{fig1}) shows the contribution from the cosmological distribution of iron sources. The shaded areas show the flux within its variation based on Eq.~(\ref{eq:combinedsig}) for a local source density $\mathcal{H}_0=10^{-5}$~Mpc${}^{-3}$ (dark gray) and $\mathcal{H}_0=10^{-6}$~Mpc${}^{-3}$ (light gray). Note  that the relative size of the error bands increases as $\mathcal{H}_0^{-1/2}$. 

The right plot of Fig.~\ref{fig3} shows the corresponding result of a proton source model with similar source parameters. The amplitude of the ensemble fluctuation is comparable to the case of iron and does not serve as a direct measure of the source composition. However, the spectral feature of the GZK-suppression is significantly different to the case of the iron model. In particular, the iron source model experiences a much steeper suppression at the upper end of the UHE CR spectrum. This poses a challenge to account for the most extreme UHE events like the $3\times10^{20}$~eV Fly's Eye event~\cite{Bird:1994mp}, if sources are too distant~\cite{Taylor:2011ta}. Candidate nearby
sources of heavy nuclei include starburst galaxies~\cite{Anchordoqui:1999cu}
and ultra-fast spinning newly-born pulsars~\cite{Blasi:2000xm,Fang:2012rx}. Our method provides a tool to distinguish ensemble fluctuations from spectral features of the the average source contribution on a statistical basis. 

Proposed future space-based CR observatories can reach integrated exposures of ${\cal O} (10^6~{\rm km}^2 \, {\rm sr} \, {\rm yr})$~\cite{Santangelo:2009zz}. This is a factor of about 100 larger than the integrated exposure reached by present ground-based air shower arrays. The statistical error of UHE CR measurements can thus be improved by a factor 10. With such a resolution the observed UHE CR spectrum can show a significant deviation from the ensemble mean due to the discreteness of close-by source contributions. For instance, the spectrum could exhibit ``spectral wiggles'' beyond the GZK suppression. Our method describes a way to quantify the bin-to-bin amplitude of these modulations for the case of a ultra-high energy cosmic ray nuclei. 
  
\section{Conclusions}\label{sec:conclusion}

In this paper we have studied the ensemble fluctuations of the mean flux and average mass number of UHE CR nuclei from the distribution of sources. We have derived an analytic expression for the relative errors which applies to the CR data at the highest energies dominated by the local sub-Gpc (low redshift) source distribution. For lower energies another analytic approximation can be derived through the use of adiabatic scaling of the source contributions. This method can be easily generalized to the case of ensemble fluctuations due to the source emission parameters such as the spectral index and the maximal energy.

As an illustration, we applied these results to a fiducial iron source model for which a homogeneous distribution of sources had previously been found to successfully reproduce the CR data within systematic uncertainties. For the case of a local source density $\mathcal{H}_0=10^{-5}$~Mpc${}^{-3}$, the resultant ensemble fluctuations of the average mass composition on top of the ensemble mean were found to exist at a tolerable level for the analysis of present generation CR  composition data below $10^{19.5}$~eV (see right panels of Fig.\ref{fig2}). 

The relative ensemble fluctuation around the mean flux increases with energy and rises above the level of 10\% at about $10^{19.8}~{\rm eV}$ (see left panels in Fig.~\ref{fig2} and left plot in Fig.\ref{fig3}). This flux variation is beyond the sensitivity of present day CR observatories as indicated by the range of the dark gray shaded bands in Fig.~\ref{fig3} in comparison to the statistical errors of CR data. The amplitude of the flux variation is similar in the case of proton models as shown in the right plot of Fig.~\ref{fig3}. However, the level of ensemble fluctuations for present generation spectral studies of the GZK suppression are potentially not ignorable if a smaller source density of $\mathcal{H}_0=10^{-6}$~Mpc${}^{-3}$ is assumed, which still remains marginally consistent with angular correlation studies assuming heavy nuclei. This is shown as the light gray shaded bands in Fig.~\ref{fig3}.

Unless the actual source densities are much larger than those considered here, next generation experiments reaching accumulated exposures of ${\cal O} (10^6~{\rm km}^2 \, {\rm sr} \, {\rm yr})$ should be sufficiently sensitive to potentially discern these fluctuations. For instance, the surface detector array of the Auger Observatory observed 25 events between $10^{19.8}$~eV to $10^{19.9}$~eV with an integrated exposure of $20,905~{\rm km}^2~{\rm sr}~{\rm yr}$~\cite{Abreu:2011pj}. With an almost fifty times larger integrated exposure of future observatories the relative Poisson error at this energy should drop below 3\% and hence would be smaller than the ensemble fluctuation, $\sqrt{\sigma_{\rm loc}^2} \agt 0.1$, even for a local density of $\mathcal{H}_0=10^{-5}$~Mpc${}^{-3}$, see Fig.~\ref{fig2}.

We have seen that ensemble fluctuations of the UHE CR spectrum increase as we decrease the minimal distance $r_{\rm min}$ to the extra-galactic source population, while keeping the local source density constant. This behavior is of course expected since the relative abundance of local sources is small $\propto4\pi r^2$ and hence more susceptible for fluctuation, but with a higher flux weight $\propto1/(4\pi r^2)$ compared to more distant populations. In fact, the ensemble variance formally diverges at {\it all} energies as we set $r_{\rm min}=0$. This is a well-known effect in studies of galactic ensemble fluctuations~\cite{Blasi:2011fi,Bernard:2012wt}. The minimal distance $r_{\rm min}$ hence serves here as a regulator. Even for a moderate minimal distance of 10Mpc and a source density of $10^{-5}$~Mpc${}^{-3}$ the ensemble fluctuation can reach large values $\sigma_{\rm loc}^2\gg1$ in the GZK region as indicated in the left column of Fig.~\ref{fig2}. This is an indication that very few or just one local source can entirely dominate the spectrum at these energies.

Some authors have considered even the extreme case that a local source like the radio galaxy Centaurus A at a distance of 3-4~Mpc can be responsible for the {\it entire} UHE CR spectrum above the ankle~\cite{Biermann:2011wf}. We have checked that for $r_{\rm min}=3$ and $\mathcal{H}_0=10^{-5}$~Mpc${}^{-3}$ the ensemble fluctuation below $10^{19}$~eV are less than 10\% and this particular ensemble realization of extra-galactic sources to appear by chance is hence unlikely within our setup. In addition, the missing anisotropy of UHE CR events challenges this interpretation. Possible caveats to this result could be large ensemble-fluctuations of intrinsic source properties that have been omitted in this study (but could easily be included) and/or strong deflections in inter-galactic magnetic fields~\cite{Anchordoqui:2011ks,Yuksel:2012ee} (see further below).

All calculations in this study consider the continuous and isotropic emission of CR sources. However, it is easy to extend the discussion for the case of episodal or anisotropic emission. If $\Delta t_{\rm src}$ is the typical timescale of the source emission and $T_{\rm exp}$ the total experimental observation time we can define an effective local source density as $\mathcal{H}'_0 = (T_{\rm exp}/\Delta t_{\rm src})\mathcal{H}_0$. This compensates the reduced continuous-equivalent emission rate $(\Delta t_{\rm src}/T_{\rm exp})Q$ of the single sources averaged over the duration of the experiment. Note that the emission rate density $\mathcal{H}Q$ that is fixed by the observed CR spectrum is independent of this rescaling. An analogous argument can be given for an anisotropic CR emitter with a preferred local (but not global) emission direction. This applies to jet-like source emission like in blazars or gamma-ray bursts. If the emission is concentrated in a cone $\Delta\Omega_{\rm src}$ we can again define a reduced effective local density $\mathcal{H}'_0 = (\Delta\Omega_{\rm src}/4\pi)\mathcal{H}_0$ that compensates the enhanced isotropic-equivalent emission rate $(4\pi/\Delta\Omega_{\rm src})Q$.

We can hence account for time-variable or anisotropic sources by rescaling the local source density. These two effects work in opposite directions. Time variability increases the number of sources contributing to the spectrum and lowers the amplitude of the ensemble fluctuation by $\sqrt{T_{\rm obs}/\Delta t_{\rm src}}$. On the other hand, anisotropy decreases the source number and increases the fluctuation by $\sqrt{4\pi/\Delta\Omega_{\rm src}}$. Note, however, that autocorrelation studies of UHE CR constrain the number of sources and limit the effective source density $\mathcal{H}'_0$. Hence, the lower limit of $10^{-6}-10^{-5}~{\rm Mpc}^{-3}$ used in this analysis does still apply.
 
In this context it is interesting to consider the effect of magnetic fields on the propagation of UHE CRs. This has been so far neglected in this study. As mentioned before, a magnetic field (regular or turbulent) has no effect on a spatially homogeneous distribution of sources. However, we break homogeneity by introducing a minimal distance $r_{\rm min}$ in our calculation. It was shown for the case of protons sources that diffusion in intergalactic magnetic fields can have a significant effect on the extra-galactic CR spectrum at lower energies. The limited distance to the closest source introduces a low energy suppression, sometimes called ``anti-GZK-cutoff''~\cite{Aloisio:2004fz}. In this case, ensemble-fluctuations at the low energies close the CR ankle can increase dramatically beyond the adiabatic value (\ref{eq:adisig}) considered in this paper.

On top of these spectral variations there is also an effect on the the effective source density $\mathcal{H}'_0$. With magnetic diffusion the time-scale $\Delta t_{\rm src}$ and emission cone $\Delta\Omega_{\rm src}$ are expected to disperse. Eventually, for strong diffusion the effective local density $\mathcal{H}'_0$ is simply the true density $\mathcal{H}_0$. Since the diffusion coefficient does in general depend on the rigidity of the CR nuclei this effect would introduce an additional energy dependence of the variation.

In conclusion, future space-based observatories with colossal exposures, ${\cal O} (10^6~{\rm km}^2 \, {\rm sr} \, {\rm yr})$, will provide the required large statistics at the high-energy end of the CR spectrum, allowing identification of ensemble fluctuation from the GZK suppression features on a statistical basis. In combination with information on the arrival-direction distribution of CRs and on the secondary fluxes of $\gamma$-rays and neutrinos these spectral features can provide a coherent picture for an {\it indirect} determination of the UHE CR nuclear composition~\cite{AhlersInPrep} and will naturally complement the current {\it direct} measurements  of extensive air shower observables.

\section*{Acknowledgments}

We would like to thank Tom Paul for valuable comments on the manuscript. MA acknowledges support by a John Bahcall Fellowship for neutrino astronomy of the Wisconsin IceCube Particle Astrophysics Center (WIPAC) and by the U.S. National Science Foundation (NSF) under grants OPP-0236449 and PHY-0236449. LAA is supported by the NSF under CAREER Grant PHY-1053663, the National Aeronautics and Space Administration (NASA) Grant 11-APRA11-0058, and the UWM Research Growth Initiative. AMT acknowledges support by a Schroedinger Fellowship at Dublin Institute for Advanced Study.  Any opinions, findings, and conclusions or recommendations expressed in this material are those of the authors and do not necessarily reflect the views of NASA or NSF.

\appendix

\section{Analytic Solution for Point-Sources}\label{sec:analytic}

The Boltzmann equations~(\ref{eq:diff0}) describe the time-evolution of an isotropic flux of CRs from a homogenous distribution of isotropic CR emitters. In the following we want to study the contribution of a single close-by ($r/H_0\ll1$) source that is continuously emitting CR nuclei. As a first step we can hence neglect the redshift loss term $\partial_E(HEY)$ in Eqs.~(\ref{eq:diff0}) and regard $Y$ as the local density of CR particles. For the case of an inhomogeneous distribution of sources we have to add a CR convection term in Eqs.~(\ref{eq:diff0}), that replaces the time-derivative for continuous emission. Since we are not interested in the direction of the source we can imagine sitting at a center of a sphere with radius $r_\star$ with surface emission rate spectrum $\mathcal{Q}_A(E)/(4\pi r_\star^2)$~\cite{Ahlers:2011sd}. The convection term in this spherically symmetric setup reduces then to $\partial_r(r^2Y)/r^2$.

\begin{figure}[t]
\centering
\includegraphics[height=2in]{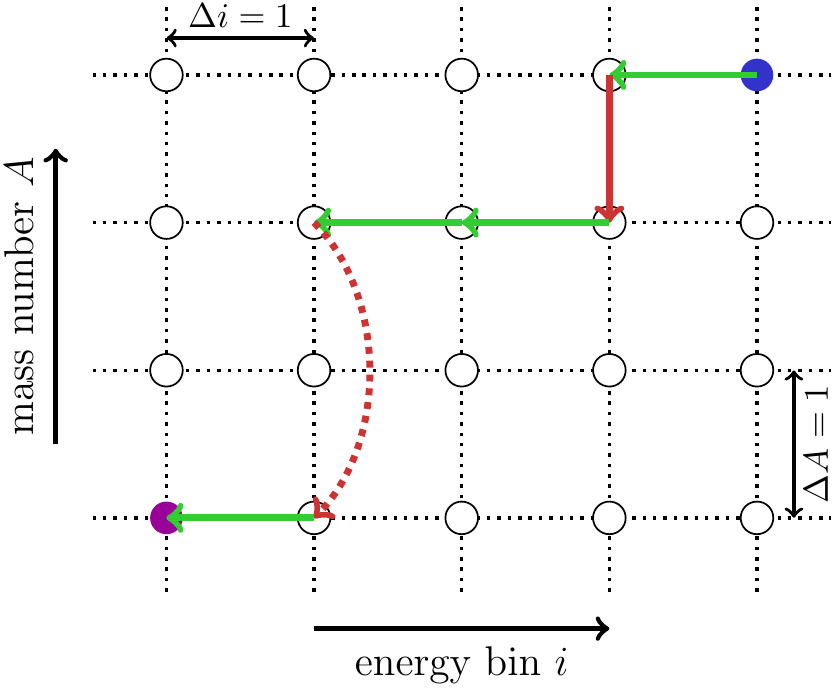}
\caption[]{A possible transition chain $\mathbf{c}$ between an initial configuration (blue dot) and a final configuration (magenta dot) including one-nucleon losses (red solid arrows), two-nucleon losses (red dotted arrows) and continuous energy loss (green horizontal arrows). For the exact analytic solution~(\ref{eq:fullsol}) all possible transition chains of this type are taken into account. (Figure adopted from Ref.~\cite{Ahlers:2010ty}.)}
\label{fig4}
\end{figure}

The secondary nuclei produced via photodisintegration carry approximately the same Lorentz factor as the initial nucleus. As already discussed in Section~\ref{sec:variation} it is hence convenient to express the energy of a nucleus with mass number $A$ as $A \en$ where $\en$ denotes the energy {\it per nucleon}.  The differential interaction rate in the set of Eqs.~(\ref{eq:diff0}) corresponding to the production of a nucleus with mass number $B$ and energy $E'$ from a nucleus with mass number $A>B$ and energy $E>E'$ can be approximated as $\gamma_{A\to B}(E,E') \simeq \Gamma_{A\to B}(E)\delta(E' - (B/A)E)$ where $\Gamma_{A\to B}$ is the partial width of the transition. 

Finally, introducing the binned CR flux $F_{A,i} \equiv \Delta \en_iA\,{{\rm d}F}_{A}(A\en_i)/{{\rm d} E}$, and corresponding emission rates, $Q_{A,i} \equiv \Delta \en_i A\mathcal{Q}_A(A\en_i)$ we can describe the point-source flux  as a solution of the compact set of equations
\begin{multline}\label{eq:diff1}
\frac{1}{r^2}\partial_r (r^2 F_{A,i}) \simeq \delta(r-r_\star)\frac{Q_{A,i}}{4\pi r^2}\\ - \sum_{B<A}\Gamma_{(A,i)\to(B,i)} F_{A,i}+ \sum_{B>A}\Gamma_{(B,i)\to(A,i)} F_{B,i}\\+\Gamma^{\rm CEL}_{A,i+1}F_{A,i+1}-\Gamma^{\rm CEL}_{A,i}F_{A,i}\,,
\end{multline}
where we define the rates:
\begin{gather}
\Gamma^{\rm CEL}_{A,i}  \equiv \frac{b_A(A\varepsilon_i)}{A\Delta \varepsilon_i}\,,
\\
\Gamma_{(A,i)\to (B,i)} \equiv \Gamma_{A\to B}(A\varepsilon_i)\,.
\end{gather}
Note that Eqs.~(\ref{eq:diff1}) holds for nuclei heavier than beryllium. We can easily compensate for the process ${}^9$Be $\to$ ${}^4$He + ${}^4$He + n of the PSB chain by re-defining $F_{A,i}' = F_{A,i}/2$ for $A=2,3,4$ and $F_{A,i}'=F_{A,i}$ for other nuclei. Similarly, the chains with nucleons as a final particle are re-weighted by the corresponding multiplicity in the case of intermediate nuclei lighter than ${}^9$Be (see Ref.~\cite{Ahlers:2010ty} for further details).

In Ref.~\cite{Ahlers:2010ty} it was shown that the general solution of Eq.~(\ref{eq:diff1}) can be written in the form of Eq.~(\ref{eq:fullsol}). The first sum in Eq.~(\ref{eq:fullsol}) runs over all possible production chains $\mathbf{c}$ of nuclei with mass $A$ in the nucleon energy bin $i$. The intermediate configurations $c_k$ of this chain are parametrized by the tuple $(B,j)$ denoting the mass number $B$ and nucleon energy bin $j$. The links of the chain correspond to the partial width $\Gamma_{c_k\to c_{k+1}}$. The dimensionless amplitudes in Eq.~(\ref{eq:fullsol}) are defined as
\begin{gather}\label{eq:A}
\mathcal{A}_k(\mathbf{c}) \equiv \prod_{l=1}^{n_c-1}\Gamma_{c_l\to c_{l+1}}\,\Big/\!\!\prod_{p=1(\neq k)}^{n_c}\left(\Gamma^{\rm tot}_{c_p}-\Gamma^{\rm tot}_{c_k}\right)\,.
\end{gather}
The total width $\Gamma^{\rm tot}_{c_k}$ is simply the total rate including photo-disintegration and CEL terms. These amplitudes satisfy the identity $\sum_k\mathcal{A}_k(\mathbf{c}) =0$ independent of the path $\mathbf{c}$ (see Refs.~\cite{Hooper:2008pm,Ahlers:2010ty}).

An example of a chain is shown in Fig.~\ref{fig4}. The elements $c_k$ of the chain correspond to intermediate configurations of mass number and energy bin on a grid. The first element $c_1$ (blue dot) corresponds to the CR particle emitted at the source; the last element $c_{n_c}$ (magenta dot) is the observed (final) configuration $(A,i)$. The example shows one-nucleon disintegration (solid red line), two-nucleon disintegration (dashed red line) and continuous energy loss (green lines).

In general, the number of possible paths is quite large and the computation of Eq.~(\ref{eq:fullsol}) numerically expensive. For the calculation in this study we have applied two more approximations to the solution (\ref{eq:fullsol}) to reduce the number of terms. Firstly, we reduced the CEL (last two terms on the r.h.s.~of Eq.~(\ref{eq:diff1})) to the effective loss term
\begin{equation}
\Gamma^{\rm CEL}_{A,i+1}F_{A,i+1}-\Gamma^{\rm CEL}_{A,i}F_{A,i}\to -\frac{b_A(A\en_i)}{A \en_i}F_{A,i}\,.
\end{equation}
This introduces a relative error of the order $\partial_E(bEF)/(bF)$, which is small in the relevant energy region of $10^{18}-10^{19}$~eV. And, secondly, we considered only the one-nucleon loss chain of the PSB approximation. It was shown in Refs.~\cite{Hooper:2008pm,Ahlers:2010ty} that this is a good approximation for CR nuclei with a large mass number, which is the focus of this analysis.

\end{document}